\documentclass[pra,showpacs,twocolumn]{revtex4}

\usepackage{graphicx}

\begin{document}
\title{How to Measure the Quantum State of Collective Atomic Spin Excitation}

\author{Koji Usami$^{1}$} \email{kusami@o.cc.titech.ac.jp}
\author{Jun-ichi Takahashi$^{2}$}
\author{Mikio Kozuma$^{1,2,3}$}  
\affiliation{$^1$PRESTO, Japan Science and Technology Agency, 4-1-8 Honcho Kawaguchi, Saitama 332-0012, Japan \\
$^2$Department of Physics, Tokyo Institute of Technology, 2-12-1 O-okayama, Meguro-ku, Tokyo 152-8550, Japan \\
$^3$CREST, Japan Science and Technology Agency, 4-1-8 Honcho Kawaguchi, Saitama 332-0012, Japan}

\date{\today }

\begin{abstract}
The spin state of an atomic ensemble can be viewed as two bosonic modes, i.e., a quantum signal mode and a $c$-numbered ``local oscillator'' mode when large numbers of spin-1/2 atoms are spin-polarized along a certain axis and collectively manipulated within the vicinity of the axis. We present a concrete procedure which determines the spin-excitation-number distribution, i.e., the diagonal elements of the density matrix in the Dicke basis for the collective spin state. By seeing the collective spin state as a statistical mixture of the inherently-entangled Dicke states, the physical picture of its multi-particle entanglement is made clear.
\end{abstract}


\maketitle

\section{Introduction}
\label{sec:Introduction}

Squeezing the spin projection noise~\cite{IBBGHMRW1993a} below the standard quantum limit~\cite{ACGT1972a} necessarily means multi-particle entanglement among elementary spins~\cite{KU1993a,KCL2005}. Such a squeezed spin state (SSS)~\cite{KU1993a} is believed to improve the precision of atomic clocks~\cite{MRKSIMW2001,OPGTVMSP2005a,CSSJ2006}, and to boost the sensitivity of atomic magnetometers~\cite{GSM2005} and atomic interferometers~\cite{OTFYK2001s,TOPKqph}. 

The SSSs have been generated in several laboratories since the pioneering experiment with cold cesium atoms~\cite{HSSP1999}. To characterize the quantum nature of the SSSs, the spin operators are usually bosonized such that $\hat{X}_{A}=\hat{J}_{y}/\sqrt{J_{x}}$ and $\hat{P}_{A}=\hat{J}_{z}/\sqrt{J_{x}}$, where $[\hat{J}_{y},\hat{J}_{z}]=iJ_{x}$ with $J_{x} \equiv \langle \hat{J}_{x} \rangle$ being $c$-numbered quantity~\cite{HSSP1999,JKP2001n} and these variances are analyzed in an analogous fashion to quantum optics~\cite{Leonhardt}. Given that in quantum optics single-mode squeezed state does not evoke the concept of multi-particle entanglement, the connection between the SSSs and their entanglement seems to be obscure at first glance.

Here we provide a different view on the collective spin state. In our framework the spin operators are considered to be made up of two bosonic modes, i.e., a quantum signal mode and a $c$-numbered ``local oscillator (LO)'' mode. This viewpoint smoothly links the quadratures and the density matrix in the inherently-entangled Dicke basis, just like the link between the quadratures and the density matrix in the inherently-nonclassical Fock basis in quantum optics~\cite{Leonhardt,SBPMM1996,LHABMS2001}. We show a concrete procedure to reconstruct the spin-excitation-number distribution, i.e., the diagonal elements of the density matrix in the Dicke basis for general collective spin states including the SSSs. The resultant statistics of spin excitations give us clear physical picture of multi-particle entanglement for collective spin states.

\section{Symmetric collective spin as two bosonic modes}
\label{Symmetric collective spin as two bosonic modes}

We start by examining the optical homodyne detection from a perspective more appropriate for grasping the concept of ``LO'' mode in a collective spin system. Let $\hat{b}_{V}$ and $\hat{b}_{H}$ denote the annihilation operators for the quantum mode and the LO mode, respectively. The dynamics of the two bosonic modes can be analyzed by angular momentum operators (Stokes operators)~\cite{YMK1986a}, which are given by 
\begin{eqnarray}
\hat{S}_{x}&=& (\hat{b}_{V}^{\dagger}\hat{b}_{V}-\hat{b}_{H}^{\dagger}\hat{b}_{H}) \nonumber \\
\hat{S}_{y}&=& (\hat{b}_{V}^{\dagger}\hat{b}_{H}+\hat{b}_{H}^{\dagger}\hat{b}_{V}) \nonumber \\
\hat{S}_{z}&=& -i(\hat{b}_{V}^{\dagger}\hat{b}_{H}-\hat{b}_{H}^{\dagger}\hat{b}_{V}).  \label{eq:S}
\end{eqnarray}
The operator $\hat{b}_{H}$ for the LO mode can be assumed to be a $c$-numbered quantity $\sqrt{n}e^{-i \varphi}$ since the number of photons excited in this mode, $\langle \hat{b}_{H}^{\dagger}\hat{b}_{H} \rangle =n$, is enormously large. The angular momentum operator $\hat{S}_{z}$ then becomes $\hat{S}_{z}=-\sqrt{2n}(\sin \varphi \ \hat{q}_{V}+ \cos \varphi \ \hat{p}_{V})$, where $\hat{q}_{V}\equiv (\hat{b}_{V}^{\dagger}+\hat{b}_{V})/\sqrt{2}$ and $\hat{p}_{V}\equiv i(\hat{b}_{V}^{\dagger}-\hat{b}_{V})/\sqrt{2}$ are the quadrature operators. The operator $\hat{S}_{z}$ is thus nothing but the observable of the homodyne detection~\cite{Leonhardt} with $\varphi$ representing the LO phase and $\sqrt{n}$ representing the mean LO amplitude.

The converse of the above argument also holds: the dynamics of a collection of $N$ spin-1/2 atoms with the symmetric collective spin operators~\cite{Dicke1954pr}, $\hat{J}_{x}$, $\hat{J}_{y}$, and $\hat{J}_{z}$, can be analyzed by two bosonic modes. The corresponding annihilation operators $\hat{a}_{\uparrow}$ and $\hat{a}_{\downarrow}$ for spin-up and spin-down modes are defined as $\hat{a}_{\uparrow}|M \rangle_{\uparrow}=\sqrt{M}|M-1\rangle_{\uparrow}$ and $\hat{a}_{\downarrow}|M \rangle_{\downarrow}=\sqrt{M}|M-1\rangle_{\downarrow}$, respectively. Since the x-axis is defined as the quantization axis, we have
\begin{eqnarray}
\hat{J}_{x}&=& \frac{1}{2}(\hat{a}_{\uparrow}^{\dagger}\hat{a}_{\uparrow}-\hat{a}_{\downarrow}^{\dagger}\hat{a}_{\downarrow}) \nonumber \\
\hat{J}_{y}&=& \frac{1}{2}(\hat{a}_{\uparrow}^{\dagger}\hat{a}_{\downarrow}+\hat{a}_{\downarrow}^{\dagger}\hat{a}_{\uparrow}) \nonumber \\
\hat{J}_{z}&=& -\frac{i}{2}(\hat{a}_{\uparrow}^{\dagger}\hat{a}_{\downarrow}-\hat{a}_{\downarrow}^{\dagger}\hat{a}_{\uparrow}).  \label{eq:J}
\end{eqnarray}
When all atoms are condensed in the spin-down mode, the state can be written as $|0 \rangle_{\uparrow}\otimes |N \rangle_{\downarrow}$. This state is called the coherent spin state (CSS)~\cite{ACGT1972a} which defines the standard quantum limit, and is the eigenstate of the operator $\hat{J}_{x}$ in Eq.~(\ref{eq:J}) with the eigenvalue $-N/2$. $|M \rangle_{\uparrow}\otimes |N-M \rangle_{\downarrow}$ then corresponds to the eigenstate of the operator $\hat{J}_{x}$ with the eigenvalue $-N/2+M$, that is, $M$ atoms excited into the spin-up mode with $N-M$ atoms remaining spin-down, and is a multi-particle entangled state except for $M=0$ and $N$. For instance, the state $|1 \rangle_{\uparrow}\otimes |N-1 \rangle_{\downarrow}$ can be described in terms of elementary spins ($|\uparrow\rangle_{i}$ and $|\downarrow\rangle_{i}$) as
\begin{eqnarray}
|1 \rangle_{\uparrow}\otimes |N-1 \rangle_{\downarrow} \nonumber &=& \frac{1}{\sqrt{N}}(|\uparrow\rangle_{1} |\downarrow\rangle_{2} \cdots |\downarrow\rangle_{N}) \nonumber \\
&+& \frac{1}{\sqrt{N}}(|\downarrow\rangle_{1} |\uparrow\rangle_{2} \cdots |\downarrow\rangle_{N})  \nonumber \\
\cdots &+& \frac{1}{\sqrt{N}}(|\downarrow\rangle_{1} |\downarrow\rangle_{2} \cdots |\uparrow\rangle_{N}) \label{eq:W}
\end{eqnarray}
and does have multi-particle entanglement~\cite{KCL2005}. Given that a collective state belongs to the symmetric subspace of the full Hilbert space (spanned by $2^{N}$ states of $N$ spin-1/2 atoms), it can be expressed by using these entangled states, namely, by the Dicke basis.

With this second quantized formalism of spin degree of freedom, the above treatment of optical homodyne detection can be translated into the language of a collective spin. For states in the vicinity of the CSS ($|0 \rangle_{\uparrow}\otimes |N \rangle_{\downarrow}$), the operator $\hat{J}_{z}$ can be modified to
\begin{equation}
\hat{J}_{z}=-\sqrt{N/2}(\sin \vartheta \ \hat{q}_{\uparrow}+ \cos \vartheta \ \hat{p}_{\uparrow}) .\label{eq:quad}
\end{equation}
This follows from the assumption that the operator $\hat{a}_{\downarrow}$ for the spin-down mode is a $c$-numbered quantity $\sqrt{N}e^{-i \vartheta}$, since the number of spin-down atoms ($\langle \hat{a}_{\downarrow}^{\dagger}\hat{a}_{\downarrow} \rangle =N$) is enormously large. Here $\hat{q}_{\uparrow}\equiv (\hat{b}_{\uparrow}^{\dagger}+\hat{b}_{\uparrow})/\sqrt{2}$ and $\hat{p}_{\uparrow}\equiv i(\hat{b}_{\uparrow}^{\dagger}-\hat{b}_{\uparrow})/\sqrt{2}$ are the collective spin analogues of the quadrature operators (the physical meaning of the quadrature distribution for the collective spin is briefly discussed in the Appendix). This approximation is similar to that for the optical homodyne detection and somewhat different from the spin quadratures $\hat{X}_{A}$ and $\hat{P}_{A}$~\cite{HSSP1999,JKP2001n} as we explicitly introduce two bosonic modes. The atoms condensed in the spin-down mode act as the ``LO" mode, collectively amplifying the tiny differences of the quadrature distributions for the spin-up atomic states. This suggests that tomographic measurements~\cite{Leonhardt} of the quantum state of the collective spin excitations, such as the squeezed spin state (SSS)~\cite{KU1993a} and the Dicke state~\cite{Dicke1954pr}, may be possible in an analogous fashion to the tomography of the squeezed state of light~\cite{SBPMM1996} and the single photon state~\cite{LHABMS2001}, respectively.

\section{Tomographic measurement of collective spin}
\label{Tomographic measurement of collective spin}

Let us consider how to measure the quadrature for a collective spin, i.e., the operator $\hat{J}_{z}$ in Eq.~(\ref{eq:quad}). In order to do this, it is required us to access only the symmetric collective spin operators~\cite{Dicke1954pr}, i.e., to validate the second quantized formalism of spin. Although the most intuitive approach may be to measure $\hat{J}_{z}$ by the Stern-Gerlach technique, the atoms respond to Stern-Gerlach magnetic field individually rather than collectively~\cite{KSJ2001}. Kuzmich \textit{et al.} showed that the atomic ensemble collectively interacts with an off-resonent optical probe and that the collective spin operator $\hat{J}_{z}$ can be measured in a quantum nondemolition (QND) way~\cite{KMJYEB1999a}. Recently, Julsgaard~\textit{et al.}~\cite{JSCFP2004n} have applied this technique to the tomography of collective spin in the context of the quantum memory experiment, in which they proved that the conjugate variables of light, i.e., the quadratures, $\hat{X}_{L}$ and $\hat{P}_{L}$, were faithfully mapped onto the atomic quadratures $\hat{P}_{A}$ and $\hat{X}_{A}$ ($\hat{p}_{\uparrow}$ and $\hat{q}_{\uparrow}$ in our language), respectively. The effective Hamiltonian describing spin QND measurement, namely dispersive vector light-shift interaction, is given by~\cite{KMJYEB1999a,GSMqph}
\begin{equation}
\hat{H}_{I}= \hbar g \hat{S}_{z}\hat{J}_{z},  \label{eq:sQND}
\end{equation}
where $g$ is the field-atom coupling coefficient, $\hat{S}_{z}$ is the Stokes operator defined by Eq.~(\ref{eq:S}), and $\hat{J}_{z}$ is the angular momentum operator defined by Eq.~(\ref{eq:J}). Note that the bosonic operators $\hat{b}_{V}$ and $\hat{b}_{H}$ in Eq.~(\ref{eq:S}) now represent the annihilation operators for horizontally and vertically polarized photons, respectively. In the Heisenberg picture, the set of Stokes operators, which is initially defined by $\mathbf{\hat{S}}(0)=\{\hat{S}_{x}(0),\hat{S}_{y}(0),\hat{S}_{z}(0)\}$, evolves into
\begin{equation}
\left[
    \begin{array}{c}
    \hat{S}_{x}(\tau) \\
    \hat{S}_{y}(\tau) \\
    \hat{S}_{z}(\tau) 
    \end{array}
\right]
\!\!\!=\!\!\!
\left[
    \begin{array}{ccc}
    \cos (g \tau \hat{J}_{z}) & -\sin (g \tau \hat{J}_{z}) & 0 \\
    \sin (g \tau \hat{J}_{z}) & \cos (g \tau \hat{J}_{z}) & 0 \\
    0 & 0 & 1
    \end{array}
\right]
\!\!\!
\left[
    \begin{array}{c}
    \hat{S}_{x}(0) \\
    \hat{S}_{y}(0) \\
    \hat{S}_{z}(0) 
    \end{array}
\right], \label{eq:opticalTE}
\end{equation}
after passage of the interaction time $\tau$. The set of angular momentum operators, $\mathbf{\hat{J}}(0)$, evolves similarly with $\mathbf{\hat{J}}$ and $\mathbf{\hat{S}}$ interchanged. Here, $\hat{J}_{z}(0)=\hat{J}_{z}(\tau)=\hat{J}_{z}$ and $\hat{S}_{z}(0)=\hat{S}_{z}(\tau)=\hat{S}_{z}$ since $\hat{J}_{z}$ and $\hat{S}_{z}$ commutes with the interaction Hamiltonian given by Eq.~(\ref{eq:sQND}). When the initial state for the optical probe is set for horizontally polarized light, $|0 \rangle_{V} \otimes |n \rangle_{H}$, which is given by the eigenstate of the operator $\hat{S}_{x}(0)$ with the eigenvalue $-n$. Then the measurement of the operator $\hat{S}_{y}(\tau)$, which can be realized by the balanced polarimeter~\cite{KMJYEB1999a}, effectively works as the quadrature measurement for a collective spin. To illustrate this concept, the operator $\hat{S}_{y}(\tau)$ in Eq.~(\ref{eq:opticalTE}) can be rewritten as $\hat{S}_{y}(\tau) \approx g \tau \hat{J}_{z}(0)\hat{S}_{x}(0)+\hat{S}_{y}(0)$ by assuming $g \tau \langle \hat{J}_{z} \rangle \ll 1$. Considering the initial condition for the optical probe, the operators $\hat{S}_{x}(0)$ and $\hat{S}_{y}(0)$ can be approximated by $\hat{S}_{x}(0)=-n$ and $\hat{S}_{y}(0)= \sqrt{n}(\hat{b}_{V}^{\dagger}(0)+\hat{b}_{V}(0))\equiv \sqrt{2n}\hat{q}_{V}(0)$, respectively. This manipulation again applies the fact that the operator $\hat{b}_{H}(0)$ is deemed to be a $c$-numbered quantity $\sqrt{n}e^{-i \varphi}$, that is, $\langle \hat{b}_{H}^{\dagger}(0)\hat{b}_{H}(0) \rangle =n \gg 1$. Here, the phase $\varphi$ between the two modes, $|0 \rangle_{V}$ and $|n \rangle_{H}$, is irrelevant and is set at $0$. We then have 
\begin{equation}
\hat{S}_{y}(\tau)\approx g \tau n \! \sqrt{N/2}(\sin \vartheta \ \hat{q}_{\uparrow}(0) + \cos \vartheta \ \hat{p}_{\uparrow}(0)) + \sqrt{2n}\hat{q}_{V}(0). \nonumber
\end{equation}
The ``LO" phase for the collective spin, $\vartheta$, corresponds to the azimuthal angle about the x-axis and can be controlled by the magnetic field along the x-axis~\cite{JKP2001n,JSCFP2004n,GSM2004s}. The observable of the balanced polarimeter, $\hat{S}_{y}(\tau)$, can thus be viewed as the amplified quadrature for a collective spin, though it includes an additional term $\sqrt{2n}\hat{q}_{V}(0)$, which represents the shot noise due to the optical probe.

To keep things simple let us normalize the operator $\hat{S}_{y}(\tau)$ as $\hat{Q} \equiv \hat{S}_{y}(\tau)/\mathcal{N}$ where $\mathcal{N}\equiv \big( (g\tau n\sqrt{N/2})^{2}+(\sqrt{2n})^{2} \big)^{1/2}$. Then we have   
\begin{equation}
\hat{Q} = \sqrt{\eta}(\sin \vartheta \ \hat{q}_{\uparrow}(0) + \cos \vartheta \ \hat{p}_{\uparrow}(0)) \!\!+\!\!\! \sqrt{1-\eta}\,\hat{q}_{V}(0)  \label{eq:BP}
\end{equation}
with $\sqrt{\eta} = g\tau n\sqrt{N/2}/\mathcal{N}$. Let $\mathrm{P}(Q,\vartheta) \equiv \langle Q,\vartheta| (|0 \rangle_{VV}\langle 0| \otimes \hat{\rho}_{\uparrow})|Q,\vartheta \rangle$ denote the histogram obtained by measurement of the operator $\hat{Q}$, where $|Q,\vartheta \rangle$ is the eigenstate of the operator $\hat{Q}$ with the eigenvalue $Q$. Here, $\rho_{\uparrow}$ represents the density matrix of the spin-up mode before field-atom interaction. The optical-noise term $\sqrt{2n}\hat{q}_{V}(0)$ prevents simple reconstruction of the density matrix $\rho_{\uparrow}$. This is essentially the same situation where detector inefficiencies and optical losses degrade the quadrature distribution for the original photonic state, necessitating loss-error compensation for reconstruction of the original state~\cite{Leonhardt,B1998a-BDPS1999a,Lvovsky2004job}. Following the argument on detection-loss induced error in the optical homodyne detection~\cite{Leonhardt}, the histogram $\mathrm{P}(Q,\vartheta)$ is given by
\begin{equation}
\mathrm{P}(Q,\vartheta)= \int_{-\infty}^{\infty}dq' \ \mathrm{P_{\uparrow}^{(\vartheta)}}(q') \ \mathrm{P_{V}}(\frac{Q-\sqrt{\eta}q'}{\sqrt{1-\eta}}),  \label{eq:Hist}
\end{equation}
with the quadrature distribution for the collective spin state
\begin{equation}
\mathrm{P_{\uparrow}^{(\vartheta)}}(q)\equiv \!_{\uparrow}\langle q,\vartheta| \hat{\rho}_{\uparrow}|q,\vartheta \rangle_{\uparrow},  \label{eq:Hist_+}
\end{equation}
and the $\vartheta$-invariant quadrature distribution for the optical probe
\begin{equation}
\mathrm{P_{V}}(q)\equiv \!_{V}\langle q|(|0 \rangle_{VV}\langle 0|)|q \rangle_{V} = (1/\sqrt{\pi})\exp(-q^{2}),,  \label{eq:Hist_V}
\end{equation}
which is the Gaussian distribution discribing the shot noise. Here $|q,\vartheta \rangle_{\uparrow}$ and $|q \rangle_{V}$ are the eigenstates of the the operators $(\sin \vartheta \ \hat{q}_{\uparrow} + \cos \vartheta \ \hat{p}_{\uparrow})$ and $\hat{q}_{V}$, respectively. Measurement of $\hat{S}_{y}(\tau)$ thus yields the smoothed quadrature distribution for the collective spin. The tomographic reconstruction of $\hat{\rho}_{\uparrow}$ can then be realized by changing the ``LO" phase $\vartheta$ via the Larmor precession with compensation for optical shot noise as demonstrated by Julsgaard~\textit{et al.}~\cite{JSCFP2004n}.

\section{Reconstruction of spin-excitation-number distribution}
\label{Reconstruction of spin-excitation-number distribution}

The tomographic reconstruction of $\hat{\rho}_{\uparrow}$ gives us an insight into the multi-particle entanglement of the spin systems when our two-mode (quantum and $c$-numbered ``LO" modes) framework is employed. The concept of the multi-particle entanglement appears becasue of the fact that the basis states, e.g., $|1 \rangle_{\uparrow}\otimes |N-1 \rangle_{\downarrow}$, are inherently entangled states as seen in Eq.~(\ref{eq:W}). The complete information of density matrix can be obtained by the similar procedure of quantum optics outlined in Refs.~\cite{Leonhardt,B1998a-BDPS1999a,Lvovsky2004job}. The reconstruction procedure so simple as one restricted to just acquiring diagonal elements of the density matrix, which might be easier to implement, gives us clear physical picture of multi-particle entanglement for collective spin states. Thus we now recapitulate a concrete procedure for determining the diagonal elements of the density matrix in the entangled Dicke basis. By physically or numerically averaging the ``LO" phase $\vartheta$ in Eq.~(\ref{eq:Hist}), we have~\cite{Leonhardt,B1998a-BDPS1999a,Lvovsky2004job}
\begin{equation}
\mathrm{P}(Q) \equiv \frac{1}{2\pi}\int^{2\pi}_{0}d\vartheta\,\mathrm{P}(Q,\vartheta)= \sum_{M=0}^{\infty}A_{M}^{(\eta)}(Q)\ \rho_{MM},  \label{eq:PV_Hist}
\end{equation}
where $\rho_{MM} \equiv \ _{\uparrow}\langle M|\hat{\rho}_{\uparrow}|M \rangle_{\uparrow}$ and 
\begin{equation}
A_{M}^{(\eta)}(Q)=\sum_{m=0}^{M}\frac{M!(1-\eta)^{M-m}\eta^{m}}{(M-m)!m!} \frac{H_{m}^{2}(Q)\exp(-Q^{2})}{\sqrt{\pi}2^{m}m!}  \label{eq:coefficientA}
\end{equation}
with $H_{m}(q)$ denoting the Hermite polynomials. Repeating $R$ times the measurement of $\hat{Q}$ affords the histogram $\{k_{\nu}\}$, where $k_{\nu}$ represents the number of events that occurred in such a way that the measured value of $Q$ belongs to the bin with width $\delta Q$ and central point $Q_{\nu}$. The probability of obtaining a specific histogram $\{k_{\nu}\}$ is given by~\cite{B1998a-BDPS1999a}
\begin{equation}
\mathcal{P}(\{k_{\nu}\}|\{\rho_{MM}\}) = R! \prod_{\nu}\frac{1}{k_{\nu}!} \mathrm{P}_{\nu}^{\ k_{\nu}},  \label{eq:Hist_D}
\end{equation}
where $\mathrm{P}_{\nu} \equiv \mathrm{P}(Q_{\nu}) \delta Q$. Equations~(\ref{eq:PV_Hist}) and (\ref{eq:Hist_D}) thus provide the relationship between the measurable histogram $\{k_{\nu}\}$ and the distribution of spin excitation number $\{\rho_{MM}\}$. 

Two obstacles are encountered in reconstructing $\{\rho_{MM}\}$ from the measured histogram $\{k_{\nu}\}$: the distribution $\{\rho_{MM}\}$ must satisfy the condition $\rho_{MM} \ge 0$ for any $M$, and the number of unknown parameters $\{\rho_{MM}\}$ is large (i.e., the same as the atom number $N$, although $\rho_{MM} \sim 0$ for $M \gg 1$). To satisfy positivity constraints for $\{\rho_{MM}\}$ and restrict the parameter space with a cutoff parameter $K$, a new set of parameters $\{\theta_{M}\}$ is defined by $\{\rho_{00}=\theta_{0}^{2}/\Theta^{(K)},\,\rho_{11}=\theta_{1}^{2}/\Theta^{(K)},\,\cdots,\,\rho_{KK}=\theta_{K}^{2}/\Theta^{(K)}\}$ with $\Theta^{(K)}=\theta_{0}^{2}+\theta_{1}^{2}+ \cdots + \theta_{K}^{2}$. To search for the best estimate of $\{\theta_{M}\}$ and choose the most appropriate cutoff parameter $K$, the maximum likelihood method with Akaike's information criterion (AIC) is employed~\cite{Akaike1974ieeeAC,UNTMN2003a}. The best estimate of $\{ \theta_{M} \}$ with the best cutoff parameter $K$ can be found by numerically minimizing the AIC:
\begin{equation}
\mathcal{A}^{(K)}(\{k_{\nu}\}|\{ \theta_{M} \})=-2 \ln \Big[ \mathcal{P}^{(K)}(\{k_{\nu}\}|\{\ \rho_{MM} \}) \Big] +2K.  \label{eq:AIC}
\end{equation}
Here, the first term denotes the log-likelihood function for the $K$-dimensional parametric model of $\mathcal{P}(\{k_{\nu}\}|\{\rho_{MM}\})$, and the second term serves to suppress $K$ and thus reduce the error due to redundant parameters~\cite{Akaike1974ieeeAC,UNTMN2003a}.

\begin{figure}
\begin{center}
\includegraphics[width=\linewidth]{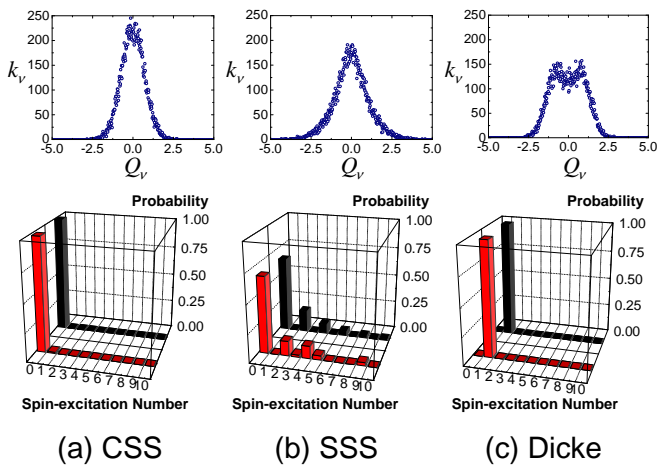}
\caption{(Color online) Simulated phase-averaged quadrature distributions $\{k_{\nu}\}$, where $k_{\nu}$ represents the number of events that occurred in such a way that the measured value of $Q$ in Eq.~(\ref{eq:BP}) belongs to the bin with width $\delta Q$ and central point $Q_{\nu}$ (upper) and estimated spin-excitation-number distributions (lower, front row) for the (a) CSS $|0\rangle_{\uparrow}$, (b) SSS $\hat{S}(\xi=1)|0\rangle_{\uparrow}$, and (c) Dicke state $|1\rangle_{\uparrow}$. The true number distributions used to simulate the quadrature distributions $\{k_{\nu}\}$ are shown in the back row of the lower figures. The resultant cutoff parameters for minimizing the AIC (Eq.~(\ref{eq:AIC})) are (a) $K=1$, (b) $K=9$, and (c) $K=1$.}
\label{fig:CStomo}
\end{center}
\end{figure}

Histograms $\{k_{\nu}\}$ for the CSS $|0\rangle_{\uparrow}$, SSS $\hat{S}(\xi=1)|0\rangle_{\uparrow}$ (where $\hat{S}(\xi)$ is the squeezing operator~\cite{Leonhardt}), and Dicke state $|1\rangle_{\uparrow}$ were generated by Monte Carlo simulation (20000 samples) with consideration of shot noise due to the optical probe. The histograms and reconstructed spin-excitation-number distributions for the three states are shown in Fig.~\ref{fig:CStomo}. In this figure, the signal-to-noise ratio is fixed at $\eta/(1-\eta)=1$, since it can be estimated separately in experiments. The result for the SSS (Fig.~\ref{fig:CStomo}~(b)) exhibits the collective-spin counterpart of the oscillatory number distribution~\cite{SBPMM1996}. The disappearance of odd-number spin excitations, such as $|1 \rangle_{\uparrow}$ and $|3 \rangle_{\uparrow}$, is a general feature of the squeezing below the standard quantum limit~\cite{SW1987n}. The phase-averaged SSS (Fig.~\ref{fig:CStomo}~(b)) can thus be expressed by a classical mixture of a separable CSS and entangled Dicke states. In this way, the statistics of spin excitations in the entangled Dicke states give us clear physical picture of multi-particle entanglement for collective spin states~\cite{Vogel2000-RV2002}. It should be emphasized that the characteristics of the small number of atoms ($\langle \hat{a}_{\uparrow}^{\dagger}\hat{a}_{\uparrow} \rangle \sim 1$) in the spin-up mode, i.e., the quantum mode, are emergent in the changes in the quadrature distributions via macroscopic number of atoms (e.g., $\langle \hat{a}_{\downarrow}^{\dagger}\hat{a}_{\downarrow} \rangle \sim 10^{11}$~\cite{GSM2004s}, or $10^{12}$~\cite{JKP2001n}) condensed in the spin-down mode, i.e., the ``LO" mode.

\section{Experimental feasibility}
\label{Experimental feasibility}

Recently the reconstruction of the CSS has been reported by Julsgaard~\textit{et al.}~\cite{JSCFP2004n} in the context of the dispersive quantum memory experiment. To observe the non-trivial spin-excitation-number distributions for e.g., the SSS and Dicke state, several schemes might be used: 

1) The aforementioned dispersive quantum memory scheme could  lead to the observations of the spin-excitation-number distributions for the SSS and Dicke state by mapping a squeezed state of light and a single photon state, respectively, onto collective spin states. 

2) Instead of the coherent media transfer between light and collactive atomic spin via dispersive quantum memory, measurement-induced non-unitary schemes can also be utilized to generate the non-classical spin state. For the SSS the preparation and observation of the spin-excitation-number distribution shwon in Fig.~\ref{fig:CStomo} (b) is readily envisaged from the experimental results~\cite{JKP2001n,GSM2004s}. To observe the Dicke state $|1\rangle_{\uparrow}$, the scheme proposed by Takahashi~\textit{et al.}~\cite{THTTIY1999a} may be used by incorporating single photon detection. The so-called DLCZ scheme~\cite{DLCZ2001n} may also be used for generating the pseudo-spin analogue of the Dicke state, and the hyperfine-dependent scalar or tensor light-shift interaction~\cite{OPGTVMSP2005a,CSSJ2006} can be used to measure $\hat{Q}$ of Eq.~(\ref{eq:BP}) instead of vector light-shift interaction (Eq.~(\ref{eq:sQND})). Also noted is the experiment done by Black~\textit{et al.}~\cite{BTV2005}. With their cavity-based magnetic storage configuration, the Dicke state generation and observation could be realized in a straightforward manner. 

3) Finally, we mention the resonant quantum memory scheme with dark-state polariton~\cite{FL2000-FL2002a}. This scheme may also fit into our reconstruction scheme when one employs the experimental geometory similar to that of Black~\textit{et al.}~\cite{BTV2005}, i.e., a ``coupling" laser beam and a ``probe" laser beam for electromagnetically induced transparency forming the magnetic storage configuration. We think that it is worth doing to investigate further the difference between the quantum states of collective atomic memony in dispersive scheme~\cite{JSCFP2004n} and resonant scheme with dark-state polariton~\cite{FL2000-FL2002a}.

\section{Conclusion}
\label{Conclusion}

We show that the quantum state of the collective spin excitation can be reconstructed from the quadrature distribution, which can be obtained by the spin quantum nondemolition measurements. The present reconstruction scheme works for any states provided the mean spin excitation in the quantum mode is sufficiently small compared to that of the ``LO" mode. Although this fact may exclude the application of our reconstruction scheme to e.g., the super-radiant Dicke state $|N/2 \rangle_{\uparrow}\otimes |N/2 \rangle_{\downarrow}$~\cite{Dicke1954pr} and the cluster state~\cite{RB2001}, the scheme may be a useful tool for investigating the multi-particle entanglement of atomic collective spin systems. Further extension to other systems such as collective nuclear spin excitation~\cite{TML2003-TIL2003-EAM2004jpsj} can readily be envisioned.

\acknowledgments
We would like to thank Daisuke~Akamatsu, Keiichirou~Akiba, Keiji~Murata, Yoshiyuki~Tsuda, Masahito~Ueda, and Go~Yusa for valuable discussions.

\appendix
\section {physical meaning of spin quadrature}

What is the physical meaning of the quadrature distribution for collective spin? We preliminarily set $\vartheta=\pi/2$ without loss of generality. The quadrature operator $\hat{q}_{\uparrow}$ appears as a result of the approximation of $\hat{J}_{z}$ in Eq.~(\ref{eq:quad}). Thus the quadrature distribution $\mathrm{P_{\uparrow}}(q)$ (Eq.~(\ref{eq:Hist_+})) is originally
\begin{equation}
\mathrm{P_{z}}(m)\equiv \langle \hat{J}_{z}=m| \hat{\rho}_{\uparrow}|\hat{J}_{z}=m \rangle.
\end{equation}
For example, $\mathrm{P_{z}}(m)$ for the CSS, $\hat{\rho}_{\uparrow}=|0 \rangle_{\uparrow \uparrow} \langle 0|$ ($|\hat{J}_{x}=-N/2 \rangle \langle \hat{J}_{x}=-N/2|$), can be written as
\begin{equation}
\mathrm{P_{z}}(m) = | \langle \hat{J}_{z}=m|\hat{J}_{x}=-\frac{N}{2} \rangle |^{2}= \Big| d^{\ \frac{N}{2}}_{m,-\frac{N}{2}}\,(\frac{\pi}{2}) \Big|^{2}
\end{equation}
with $d^{N/2}_{m,-N/2}\,(\pi/2)$ being the nontrivial part of the rotaion matrix element for a spin-N/2 system~\cite{Zare}. From the explicit expression~\cite{Zare}, 
\begin{equation}
d^{\ \frac{N}{2}}_{m,-\frac{N}{2}}\,(\frac{\pi}{2}) = \frac{N!}{\sqrt{(\frac{N}{2}+m)!(\frac{N}{2}-m)!}} (\frac{1}{2})^{\frac{N}{2}}, 
\end{equation}
and the de Moivre-Laplace theorem, we have a Gaussian distribution, given by 
\begin{equation}
\mathrm{P_{z}}(m) \approx \frac{1}{\sqrt{\pi \frac{N}{2}}} \exp[-\frac{m^{2}}{\frac{N}{2}}], 
\end{equation}
as the large-N limit. This form is equivalent to the $\sqrt{N/2}$-folded quadrature distribution for the vacuum state, which is given by $\mathrm{P_{\uparrow}}(q)= \!_{\uparrow}\langle q|(|0 \rangle_{\uparrow \uparrow}\langle 0|)|q \rangle_{\uparrow}$. 

As another example, the quadrature distribution for the Dicke state, $\hat{\rho}_{\uparrow}=|1 \rangle_{\uparrow \uparrow} \langle 1|$ ($|\hat{J}_{x}=-(N/2)+1 \rangle \langle \hat{J}_{x}=-(N/2)+1|$), can be written as 
\begin{equation}
\mathrm{P_{z}}(m) = | \langle \hat{J}_{z}=m|\hat{J}_{x}=-\frac{N}{2}+1 \rangle |^{2} =  \Big| d^{\frac{N}{2}}_{m,-\frac{N}{2}+1}\,(\frac{\pi}{2}) \Big|^{2},
\end{equation}
and its large-N limit becomes 
\begin{equation}
\mathrm{P_{z}}(m) \approx \frac{1}{\sqrt{\pi \frac{N}{2}}} \big( -2 \frac{m^{2}}{\frac{N}{2}} \big) \exp[-\frac{m^{2}}{\frac{N}{2}}].
\end{equation}
This form is again equivalent to the $\sqrt{N/2}$-folded quadrature distribution for the state of a single spin excitation, that is, $\mathrm{P_{\uparrow}}(q)= \!_{\uparrow}\langle q|(|1 \rangle_{\uparrow \uparrow}\langle 1|)|q \rangle_{\uparrow}$. Thus the $\sqrt{N/2}$-folded spin quadrature $\mathrm{P_{\uparrow}}(q)$ is just the probability distribution of collective spin projecting onto the z-axis, i.e., $\mathrm{P_{z}}(m)$, in the large-N limit.

\end{document}